\documentclass[prl,aps,epsfig,twocolumn,showpacs,amsmath,amssymb]{revtex4} 
 
\usepackage{graphicx} 
\usepackage[dvips]{epsfig}
\usepackage{float}

\newcommand{\be}{\begin{equation}}
\newcommand{\ee}{\end{equation}}

\begin{document} 

\title{Exact numerical simulations of  a one-dimensional, trapped
Bose gas}
\author{Bernd Schmidt}
\affiliation{Fachbereich Physik, Technische Universit\"{a}t 
Kaiserslautern, D-67663 Kaiserslautern, 
Germany} 
\author{Michael Fleischhauer}
\affiliation{Fachbereich Physik, Technische Universit\"{a}t 
Kaiserslautern, D-67663 Kaiserslautern, 
Germany}

\date{\today} 
\begin{abstract}
We analyze the ground-state and low-temperature properties of a one-dimensional
Bose gas in a harmonic trapping potential using the numerical density matrix 
renormalization group. 
Calculations cover the whole range from the Bogoliubov limit of weak interactions
to the Tonks-Girardeau limit. Local quantities such as density and local three-body correlations are calculated and shown to agree very well with analytic predictions 
within a local density approximation. The transition between temperature
dominated to quantum dominated correlation is determined and it is shown that 
despite the presence of the harmonic trapping potential first-order correlations 
display over a large range the algebraic decay of a harmonic fluid with a 
Luttinger parameter determined by the density at the trap center. 
\end{abstract}
\maketitle 


Stimulated by the recent experimental progress in generating ultracold
trapped quantum gases in one dimension \cite{Moritz-PRL-2003,Stoeferle-PRL-2004,Tolra-PRL-2004,Paredes-Nature-2004,Kinoshita-Science-2004,Kinoshita-Nature-2006} there is a growing interest in correlation 
properties
of these systems. The physics of one-dimensional quantum gases is distinct from that in higher dimensions
as it is dominated by quantum fluctuations. In a homogeneous
system of Bosons there is no long-range order even at $T=0$; correlations decay as a power-law
due to zero-point phase fluctuations. At any finite $T$ there is an asymptotic exponential decay.
The most peculiar property of interacting Bosons in 1D is the transition to 
the fermion-like Tonks-Girardeau gas \cite{Tonks,Girardeau} for small densities
or large interactions. The transition is characterized by a single effective interaction parameter, the Tonks parameter $\gamma$, where small values correspond to the weak-interaction or Bogoliubov limit 
and large values to the Tonks-Girardeau limit.  The homogeneous gas is exactly solvable
by Bethe ansatz for $T=0$ \cite{Lieb1963} and finite $T$ \cite{Yang-Yang}. 
Correlation properties can however not easily be extracted from the Lieb-Liniger solution 
\cite{Korepin-book-1993} and  require in general numerical techniques such as
Monte-Carlo simulations \cite{Astrakharchik-PRA-2003}. Approximate analytic expressions 
can be obtained only for small distances 
\cite{Olshanii-PRL-2003} or within the harmonic-fluid approach \cite{Haldane-PRL-1981,Monien-PRA-1998}. 

In the presence of a trap potential $V(x)$ 
integrability is lost. In order to nevertheless calculate local properties Bethe-ansatz solutions for the homogeneous gas can be employed together with
a  local density approximation (LDA) and the Hellman-Feynman theorem \cite{Kheruntsyan-PRL-2003}.
Recently we have used stochastic simulation techniques to calculate the density profile and
first-order correlations of a 1D Bose gas in a harmonic trap \cite{Schmidt-PRA-2005}. The stochastic
simulations were however limited to temperatures larger than the trap
energy $k_B T\approx \hbar\omega$ and thus did not allow to go deeper into the quantum  regime. In the present paper we develop an alternative numerical approach
based on the density-matrix renormalization group (DMRG) \cite{Schollwoeck-RMP-2005,Kollath-PRA-2004} which leads 
to results with much higher precision for temperatures form zero to $\hbar\omega$. 

We consider a one-dimensional Bose gas with delta interaction in a (harmonic) 
trapping potential $V(x)$
\begin{align}
\label{Ham}
\hat H=\int dx\hat\Psi^\dagger(x)\left[-\frac{1}{2}\frac{\partial^2}{\partial x^2}+V(x)+\frac{g}{2}\hat\Psi^\dagger(x)\hat\Psi(x)\right]\hat\Psi(x),
\end{align}
where we have used oscillator units, i.e. $\hbar=m=1$. $g$ is the 1D interaction strength proportional
to the s-wave scattering length in one dimension $a_{\rm 1D}$ 
\cite{Olshani98}. 

In the absence of an external trapping potential the Hamiltonian (\ref{Ham}) is integrable in the thermodynamic
limit, i.e. it has an infinite number of constants of motion. The ground-state solution for $T=0$ which can be obtained by Bethe
ansatz \cite{Lieb1963} shows that the 1D Bose-gas is fully characterized by one parameter $\gamma={g}/{\rho}$, the so-called Tonks parameter. Here $\rho$  the density of the gas. 
The Bethe ansatz leads to the so called Lieb equation
\begin{eqnarray*}
\sigma(k)-\frac{1}{2\pi}\int^1_{-1}\!\! dq\, \frac{2\lambda\sigma(q)}{\lambda^2+(k-q)^2}=\frac{1}{2\pi},
\end{eqnarray*}
where $\lambda$ is an implicit function of $\gamma$:
$ \lambda=\gamma\int^1_{-1}dk\,\, \sigma(k)$.
All local properties of the gas can be expressed in terms of the (even) moments of
$\sigma(k)$
\begin{eqnarray*}
\epsilon_m(\gamma)=\left(\frac{\gamma}{\lambda}\right)^{m+1}\int^1_{-1}dk\,\, k^m\, \sigma(k),\quad m=2,4,\dots\quad.
\end{eqnarray*}
E.g. the equation of state reads $\mu=\mu(\rho,g)=g^2\, f(\gamma)$
\begin{equation}
f(\gamma) =\frac{3\epsilon_2(\gamma)-\gamma\epsilon_2'(\gamma)}{2\gamma^2}.\label{eq:f}
\end{equation}

Integrability is no longer given when a (harmonic) trapping potential $V(x)$ is taken
into account. An often used approximation to nevertheless describe the local properties 
in the inhomogeneous case is the local density
approximation (LDA). The LDA assumes that the homogeneous solution holds with the chemical
potential $\mu$ replaced by an effective, local chemical potential $\mu_{\rm eff}(x)=\mu-V(x)$. As long as
the characteristic length of changes is small compared to the healing length the LDA is
believed to work well. Within this approximation one finds e.g. for the density of the gas:
\begin{align}
\rho(x)=\frac{g}{f^{-1}\left(\frac{\mu_{\rm eff}(x)}{g^2}\right)}
\label{eq:density}
\end{align}
where $f^{-1}$ is the inverse function of eq.(\ref{eq:f}).

In order to develop an in principle exact numerical algorithm we here want to 
employ powerful real-space renormalization 
methods such as the DMRG \cite{Schollwoeck-RMP-2005,Kollath-PRA-2004}. To this end it is necessary 
to map the continuous to a lattice model.
As shown in \cite{Cazalilla03,Schmidt-PRA-2005} this can be done in a 
consistent way by introducing 
an equidistant grid $x_j=j\Delta x$, $j\in\mathbb{Z}$, which amounts to 
replacing the field operator $\hat\Psi(x_j)$  by $\hat a_j/\sqrt{\Delta x}$, where $\hat a_i$ is a bosonic annihilation operator. Integrals are
replaced by their corresponding sums and the second derivative in the kinetic energy
term can be safely approximated by the difference quotient $\frac{\partial^2}{\partial x^2}\hat\Psi(x_j)\approx
(\hat\Psi(x_{j+1})+\hat\Psi(x_{j-1})-2\hat\Psi(x_j))/\Delta x^2$.
This leads to the Bose-Hubbard Hamiltonian
\begin{align}
\label{Ham_BH}
\hat H=\sum_i\left[-J(\hat a_i^\dagger \hat a_{i-1}+\hat a_i^\dagger \hat a_{i+1})+D_i\hat a_i^\dagger \hat a_i+\frac{U}{2}\hat a_i^\dagger{}^2 \hat a_i^2\right],
\end{align}
where  $J={1}/{2\Delta x^2}$, $D_i={1}/{\Delta x^2}+V(x_i)-\mu$ and $U={g}/{\Delta x}$. 
Expressing the scaled hopping in terms of the Tonks parameter
at the trap center, 
${J}/{U} = ({\gamma \rho(0) \Delta x})^{-1}$ and taking 
into account that 
$a_{\rm 1D}\ll \Delta x\ll \rho(0)^{-1}$, the
1D gas corresponds to a compressible phase of the BHM with negative
effective chemical potential
approaching the line $\mu_{\rm BH}/U=-2 J/U$. In the limit of vanishing average particle number per site, 
$\rho\Delta x\to 0$, Hamiltonian (\ref{Ham}) and (\ref{Ham_BH}) 
become equivalent.

\begin{figure}[htb] 
\begin{center} 
\includegraphics[bb=0 0 25.5cm 19cm,clip=true,width=0.42\textwidth]{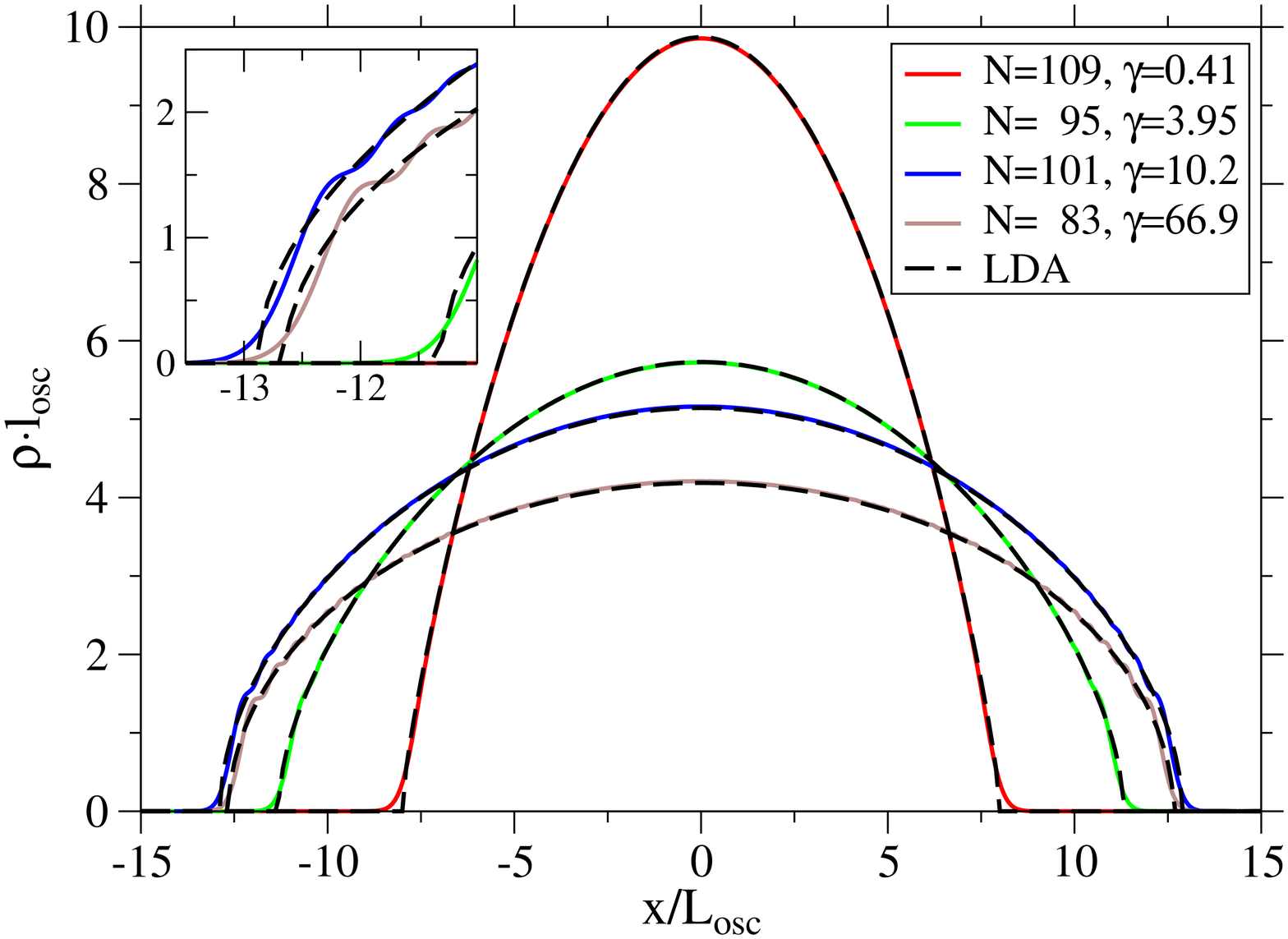}
\caption{Density of the 1D bosonic gas in a trap at $T=0$. The solid lines are the DMRG results and the dashed lines are the Lieb-Liniger prediction in local density approximation. Increasing values of $\gamma$
correspond to decreasing densities at the trap center.
Insert shows details at the edge of the density distribution.
}
\label{densitypics} 
\end{center} 
\end{figure}

The numerical DMRG calculations of the density profile, shown in Fig.(\ref{densitypics}), for Tonks parameters
$\gamma$ ranging from $0.4$ to about $70$  show excellent agreement with the Lieb-Liniger result with LDA (\ref{eq:density}) apart
from a very small region at the edges and the barely visible Friedel-type oscillations, which result
from the finite number of particles.
One recognizes the typical change of the density profile from an inverted parabola in the
Bogoliubov regime $\gamma\ll 1$ to the square root of a parabola in the Tonks-Girardeau
limit $\gamma\gg 1$ \cite{Olshani98}.

\begin{figure}[htb]
\includegraphics[bb=0 0 25.5cm 19cm,clip=true,width=0.42\textwidth]{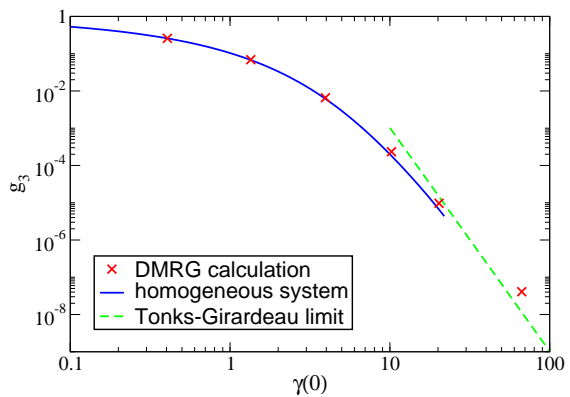}
\caption{Local 
third-order correlations as function of
Tonks parameter at the trap center (red crosses) compared to 
prediction from Lieb-Liniger theory with
local density approximation (solid line) and 
Tonks-Girardeaux limit (dashed line).}
\label{g3}
\end{figure}

An important consequence of the Fermion-like behavior of Bosons in the Tonks
limit $\gamma\gg 1$ is a dramatic reduction of the loss
rate due to inelastic three-body collisions \cite{Tolra-PRL-2004}. The rate is proportional to the local 
three particle correlation
%
%
$ g_3(x)=\langle\hat \Psi^\dagger{}^3(x)\hat\Psi^3(x)\rangle/\rho(x)^3,$
%
%
and determines the stability of the Bose gas.
Making use of the Hellman-Feynman theorem and the constants
of motion of the homogeneous gas Cheianov
\cite{Cheianov2005} has found
\begin{eqnarray}
g_3 &=&\frac{3}{2\gamma}\epsilon_4'-\frac{5\epsilon_4}{\gamma^2}+\label{eq:g3-Cheianov}\\
&&+\left(1+\frac{\gamma}{2}\right)\epsilon_2'-2\frac{\epsilon_2}{\gamma}-\frac{3\epsilon_2\epsilon_2'}{\gamma}+\frac{9\epsilon_2^2}{\gamma^2}.\nonumber
\end{eqnarray}
Fig. \ref{g3} shows a comparison between the numerical data for $g_3(0)$ at the trap center
with eq.(\ref{eq:g3-Cheianov}) and the asymptotic expression in the Tonks-Girardeau limit
with  $\gamma$ taken at the trap center $\gamma(0)=g/\rho(0)$. One recognizes again
excellent agreement except for a small deviation for very large $\gamma$, where the numerics
is however susceptible to errors due to the smallness of $g_3$. 
%
%
%
%

\begin{figure}[htb]
\includegraphics[bb=0 0 25.5cm 19cm,clip=true,width=0.42\textwidth]{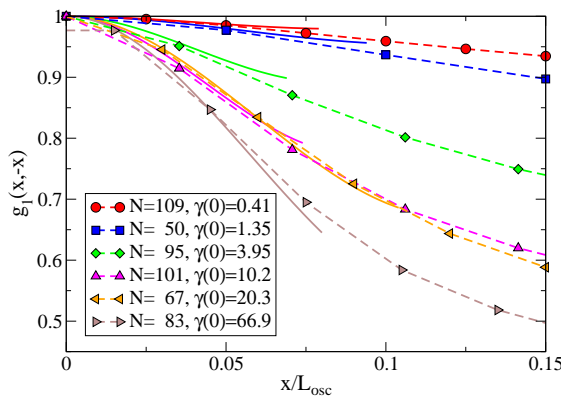}
\caption{First order correlations (dashed lines) compared to 
analytic short-distance expansion (solid lines) for a
homogeneous gas with $\gamma$ taken at the trap center. Values of $\gamma$ increase from top to
bottom curve.}
\label{fig:g1-short}
\end{figure}

In contrast to local quantities, such as the moments of the number density, information
about spatial correlations of the homogeneous 1D Bose gas such as
%
%
$g_1(x_1,x_2)=\langle\hat\Psi^\dagger(x_1)\hat\Psi(x_2)\rangle/\sqrt{\rho(x_1)\rho(x_2)}$
%
%
cannot straight-forwardly be obtained from the Lieb-Liniger and Yang-Yang theories. 
Making use of the Hellmann-Feynman theorem and the asymptotic
properties of the Lieb-Liniger wavefunction for large momenta, Olshanii and
Dunjko derived the lowest-oder terms of the Taylor expansion of $g_1$ around $x_1-x_2=0$
\cite{Olshanii-PRL-2003}
\begin{eqnarray}
g_1(x_1,x_2) &=& 1-\frac{1}{2}\Bigl(\epsilon_2(\gamma)-\gamma\epsilon_2^\prime (\gamma)\Bigr) \rho^2 x^2
\nonumber\\
&& +\frac{1}{12} \gamma^2 \epsilon_2^\prime(\gamma) \rho^3 |x|^3 +\cdots, \label{eq:g1-short}
\end{eqnarray}
with $x=x_1-x_2$. 
In the presence of a trapping potential the Tonks parameter becomes space dependent
$\gamma\rightarrow \gamma(x)$. 
Short-range correlations are however expected not to depend on the confining 
potential. Fig.\ref{fig:g1-short} shows a comparison  between $g_1$ obtained from
eq.(\ref{eq:g1-short}) and numerical results for different Tonks parameters at the
trap center. Taking into account that a high resolution of the short-distance behavior is
numerically very difficult the agreement is rather good.


The long-range or low-momentum behavior of the correlations can be obtained
from a quantum hydrodynamic approach \cite{Haldane-PRL-1981,Monien-PRA-1998} in which long-wave
properties of the 1D fluid are described in terms of two conjugate variables, the
local density fluctuations $\delta \rho$ and the phase $\phi$: $\hat \Psi(x)=\sqrt{\rho(x)}
{\rm e}^{-i\phi(x)}$. The equations of motion for $\delta\rho$ and $\phi$ follow
from the effective Hamiltonian \cite{Cazalilla03,Gangardt-PRL-2003}
\begin{equation}
H=\int\frac{{\rm d}x}{2\pi}\, \Bigl[v_N\bigl(\pi \delta\rho(x)\bigr)^2 +v_J\bigl(\partial_x\phi(x)\bigr)^2\Bigr].
\end{equation}
Here  $v_N=
(\pi)^{-1} \partial\mu/\partial \rho$ and $v_J=\pi\rho$. 

In the homogeneous case one finds that the leading-order
term in the asymptotic of the first order correlation at temperature $T$ is given by \cite{Cazalilla03}
\begin{equation}
g_1(x_1,x_2) \approx \left(\frac{K/L_T}{\rho\sinh\left(\frac{\pi|x_1-x_2|}{L_T}\right)}\right)^{1/2 K}
\label{eq:Luttinger}
\end{equation}
where $K=\sqrt{v_J/v_N}$ is the so-called Luttinger parameter and $L_T$ is the thermal
correlation length $L_T= \pi\rho/K T$, where we have set $k_B=1$.
One recognizes that for $T=0$ correlations decay asymptotically as a power-law with exponent
$1/2 K$, while for finite $T$ there is an intermediate power-law behavior turning into an exponential
decay for $|x_1-x_2| \ge L_T$. For $T=0$ the exponent $1/2K$ is given by 
\begin{equation}
\frac{1}{2K}= \frac{1}{2}\sqrt{-\frac{\gamma^3 f'(\gamma)}{\pi^2}}.\label{eq:alpha}
\end{equation}
%
%
\begin{figure}[htb] 
\begin{center} 
\includegraphics[bb=0 0 25.5cm 19cm,clip=true,width=0.42\textwidth]{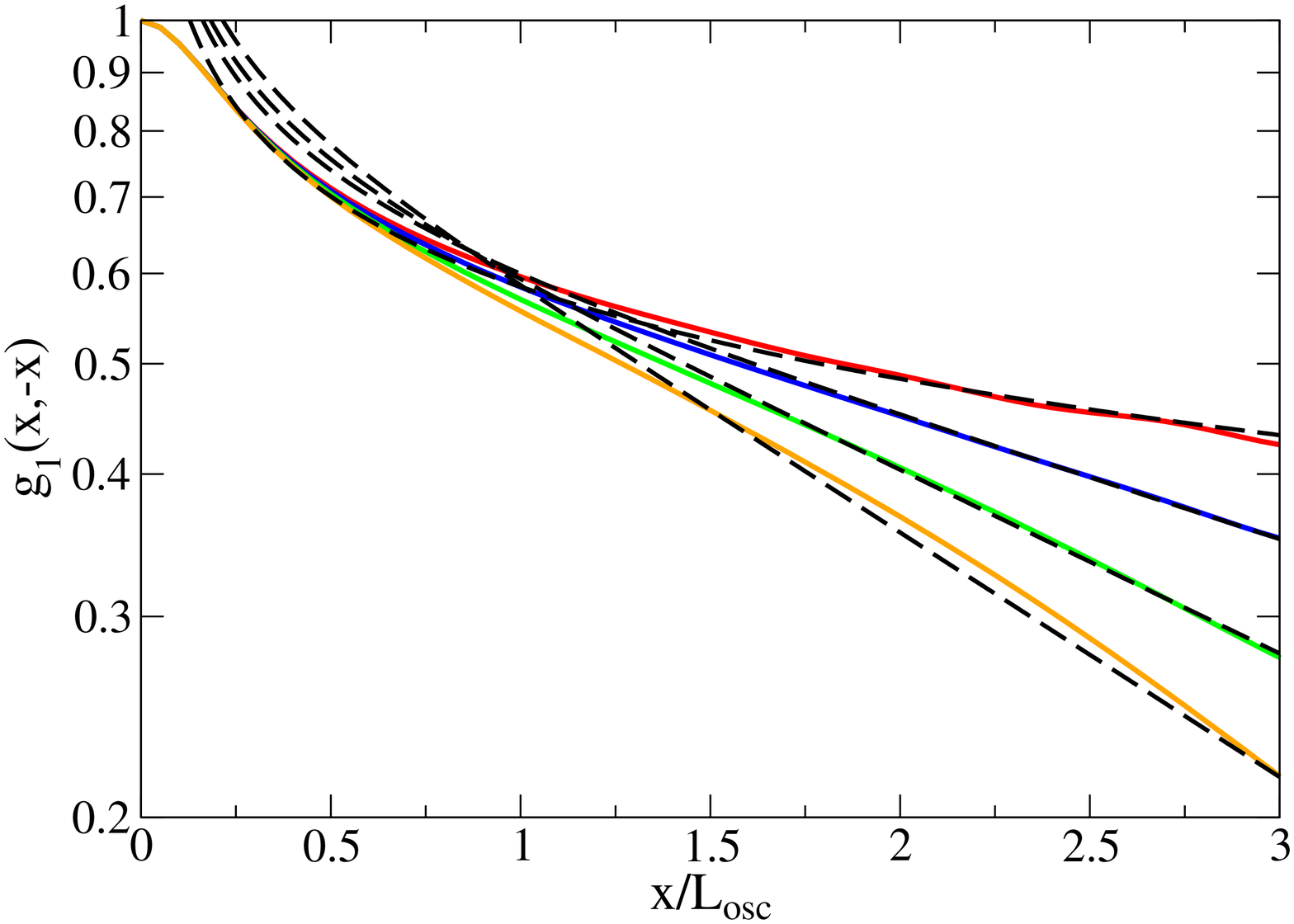}
\includegraphics[bb=0 0 25.5cm 19cm,clip=true,width=0.42\textwidth]{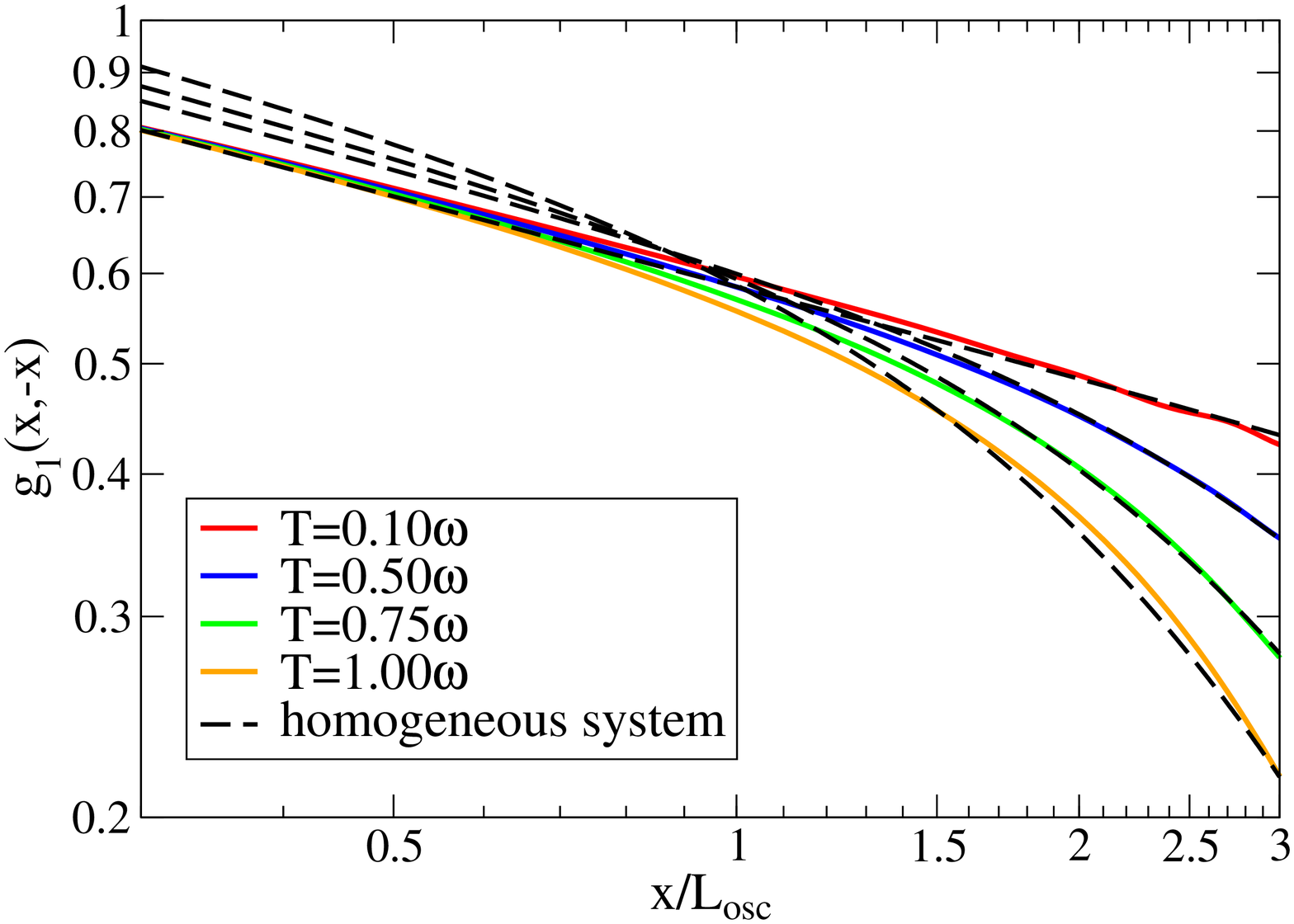}
\caption{First order correlations in the temperature regime between exponential and algebraic decay. 
{\it top:} semi-logarithmic plot, {\it bottom:} double-logarithmic. Solid curves are DMRG
calculations in the trap, dashed lines are harmonic fluid predictions for a homogeneous gas
with $\gamma$ taken at the trap center. Transition
from thermal (exponential decay) to quantum dominated correlations (algebraic decay) at $T\ll \omega$ is apparent. The parameters are: $\gamma=3.95$, $N=12$. Increasing temperature values correspond to
lower curves for large $x$.}
\label{fig:T-trans}
\end{center}
\end{figure}
%

\begin{figure}[htb]
\includegraphics[bb=0 0 25.5cm 19cm,clip=true,width=0.42\textwidth]{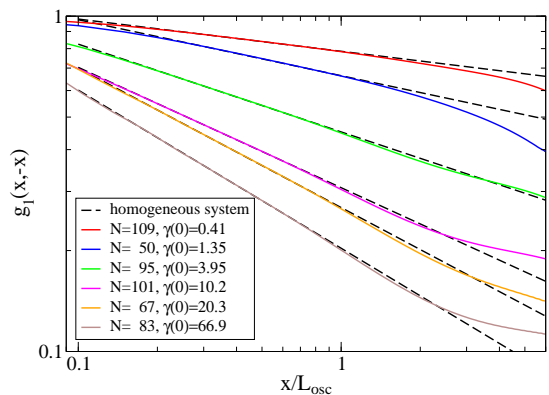}
\caption{Logarithmic plot of first-order correlations for $T=0$ and various interaction strengths (dots). The dashed lines show power-law prediction from the harmonic fluid approach with a Luttinger parameter determined by the density at the trap center. $\gamma$ increases from top to bottom curves.}
\label{fig:g1logg}
\end{figure}

In Fig.\ref{fig:T-trans} we have plotted the first-order coherence $g_1(x,-x)$ for symmetric
positions with respect to the trap center for $\gamma=3.95$ and different temperatures.
For comparison the harmonic-fluid results for the homogeneous case, eq.(\ref{eq:Luttinger}), are 
also shown with $K$ and $\rho$ taken at the trap center and for $T=0$. (The change of $K$ and
$\rho$ with $T$ has little effect). One recognizes two things: First of all
the transition from an exponential to a power-law decay happens around $T=0.1 \omega$. 
Secondly the correlations are rather well described
by the homogeneous solution (\ref{eq:Luttinger}).  A similar observation can be made at $T=0$.
Fig. \ref{fig:g1logg} shows the DMRG results for $g_1(x,-x)$ for different interaction strength.
The straight lines show the harmonic fluid predictions for the {\it homogeneous} case. Again a rather
good agreement is found for $x\le 3 L_{\rm osc}$, which on first glance is
rather surprising since the density of the gas is space dependent. The agreement is less surprising if one
notes that the local Tonks parameter $\gamma(x)$ and thus the local Luttinger parameter
$K(x)$ are almost constant within this distance range. Furthermore replacing $\rho$ in the
denominator of eq.(\ref{eq:Luttinger}) by $\rho(x)$ and expanding in a power series
one finds
%
%
%
%
that even in the Tonks limit where $K\to 1$ the corrections are small for positions sufficiently far away
from the edges of the density distribution. 

In summary we have developed a numerical scheme based on the density-matrix renormalization
group that allows to calculate local properties as well as correlations of
a 1D Bose gas in a trapping potential for temperatures up to the oscillator frequency. For local quantities
such as the density or the local three-body correlation we found excellent agreement with the
predictions from the Lieb-Liniger and Yang-Yang theories with local density approximation. Deviations
from LDA are found only in the immediate vicinity of the edges of the gas or for smaller particle
numbers where finite size effects come into play. We have shown that first-order correlations
for positions well within the gas are well described by the homogeneous theory with parameters
taken at the trap center. In particular the transition from a thermal dominated regime of exponential
decay to a power law decay of correlations was shown, with exponents as predicted by
the harmonic fluid approach in the homogeneous case for parameters taken at the trap center. 

The financial support through the DFG priority program "Ultracold quantum gases" is gratefully
acknowledged.


\end{document}